\documentstyle[preprint,prd,aps,eqsecnum]{revtex}
\begin{document}
 
\draft
 
\title{{Rotating matter in general relativity --
        stationary state I.}\thanks{Work
supported by OTKA research grant no. T17176}}
 \author{Mattias Marklund\thanks{e-mail:
mattias.marklund@physics.umu.se}}
\address{Department of Plasma Physics,
Ume{\aa} University, S-901 87 Ume{\aa}, Sweden}
\author{Zolt{\'a}n
Perj{\'e}s\thanks{e-mail: perjes@rmk530.rmki.kfki.hu}}
\address{KFKI Research Institute for Particle
and Nuclear Physics,\\Budapest 114, P.O.Box 49, H-1525
Hungary}
 
\date{\today}
 
\maketitle

\begin{abstract}
  Stationary rotating matter configurations in general relativity are
considered. A formalism for general stationary space times is developed.
Axisymmetric systems are discussed by the use of a nonholonomic and
nonrigid frame in the three-space of the time-like Killing trajectories.
Two symmetric and trace-free tensors are constructed. They characterize
a class of matter states in which both the interior
Schwarzschild and the Kerr solution are contained.
Consistency relations for this class of perfect fluids are
derived. Incompressible fluids characterized by these tensors are
investigated, and one differentially rotating solution is found.
\end{abstract}
 
 
\section{Introduction}
   In the past decades, the theory of empty space-times in general
relativity has been enriched by sophisticated results. Our understanding
of interior metrics with matter sources is apparently lagging behind:
as though time has stopped at Schwarzschild's spherically symmetric
solutions. The treat of technologies available for this research shows a
corresponding disparity. Spherical symmetry is a quite
straightforward problem to work with. Beyond, vacuum metrics have been
treated with success by coordinate, tetrad and triad
approaches, as well as the inverse scattering techniques \cite{bible}.
Regrettably, these approaches are less than adequate for successfully
treating space-times with matter. The list of known differentially
rotating
fluid metrics is surprisingly short, consisting probably of nothing else
but the unmatched perfect fluid of Chinea \cite{Chinea2}, obtained
by a tetrad method \cite{Chinea1}.
The list of matched rotating fluids is equally short;
there is the rigidly rotating disk of dust of
Meinel and Neugebauer \cite{dust} found by the inverse scattering
method.
The difficulty in finding rotating equilibrium solutions
is present already in Newtonian theory, and general relativity
should not be expected an easier field.
 
In this paper, we develop a new technology for stationary space-times
with matter sources. In Sec. II, we assume the existence of only one
Killing vector. We give Einstein's field equations with matter in a form
referring to the three-manifold ${\cal S}$ of the Killing trajectories.
In the stationary case considered here, the Killing vector is
time-like and the
metric $g$ of ${\cal S}$ is positive definite. However, the formalism
is valid independently of the signature.
 
  In Sec. III, we consider axisymmetric systems. We construct two
symmetric and traceless tensors which
are related both to the Cotton-York tensor and to the Simon tensor.
Conformally flat manifolds have the special property that the
Cotton-York tensor vanishes. This property is not shared by some
important empty space-times such as the Kerr metric. However, the
Kerr metric has a vanishing Simon tensor. We have these particular
situations in mind when we construct our tensors for space-times with
matter sources. In doing so, we are guided by two principles. First, we
want the two tensors to be uniquely determined by the Cotton-York and
the Simon tensors in the limit of some known solutions of the
gravitational equations. Thus we include some new terms in the tensors
which vanish in the absence of matter in the state of differential
rotation.
 
   Our second guiding principle stems from the theory of rotating
matter. In Sec. \ref{sec:ComplexBasis}, we initiate a new
method which synthesizes the coordinate
approach known as 'Ernst coordinates' \cite{Zol2} and the triad methods.
The Ernst coordinate approach fails in the presence of a differential
rotation of the medium for the simple reason that the Ernst potential
then does not exist. The complex 1-form, represented by the
differential of the Ernst coordinate in the vacuum, can still be
introduced although it ceases to be exact. Thus
we can build up a triad in the generic (differentially rotating) case
when the 1-form
and its complex conjugate are linearly independent. Hence we follow the
vacuum Ernst coordinate approach by setting up a complex triad in
the three space such that the directional derivatives reduce to the
partial derivatives with respect to the Ernst coordinates in the limit
of vanishing matter.
 
   In Sec. \ref{sec:ZeroSimon} we find that the coefficients of the
extra terms are determined by the requirement that
the complex invariant ${\alpha}$ be an analytic function of a complex
variable. In Sec. IV we apply our method for rotating perfect fluids.

\section{One time-like Killing vector}

  The metric of a stationary space-time with time-like
Killing vector $K=\partial/\partial t$ has the form
\begin{equation}
  {\rm d} s^2=r({\rm d} t+{\omega}_i{\rm d} x^i)^2 -
  r^{-1}{\rm d}{\ell}^2 \qquad
  (i,j,... = 1,2,3),
\end{equation}
where $r=K\cdot K$ is the length square of the Killing vector,
${\omega}_i{\rm d} x^i$ a 1-form, and
${\rm d}{\ell}^2=g_{ij} {\rm d} x^i{\rm d} x^j$ the metric in the 
3-space ${\cal S}$
of Killing trajectories. The Ricci tensor may be decomposed as 
\cite{Zol1}
\begin{mathletters}\label{eq:fieldeqs}
  \begin{eqnarray}
   -r^{-2}R^{(4)}_{oo}&=&{G^i}_{;i}+(\bar G_i-G_i)G^i ,
   \label{eq:div} \\
   -i{\epsilon_{ijk}} r^{-2}R^{(4)k}_o&=&G_{i;j} - 
   G_{j;i}+G_i\bar G_j-\bar G_i
   G_j , \label{eq:com}\\
   r^{-2}\left[g_{ik}g_{j\ell}R^{(4)k\ell} -
   g_{ij}R^{(4)}_{oo}\right]&=&R_{ij}+G_i\bar
   G_j+\bar G_i G_j , \label{eq:ricdec}
  \end{eqnarray}
\end{mathletters}
where the subscripts $o$ denote contractions with $K$.
The complex 1-form $G$ over the 3-space is defined as
\begin{equation}\label{eq:G}
  G{\stackrel{\rm def}{=}}\frac{{\rm d} r+ir^2{\rm d}{\omega}^*}{2r} ,
\end{equation}
with ${\rm d}{\omega}^*={\epsilon_{ijk}}{\omega}^{j;k}{\rm d} x^i$ 
the dual of the exterior derivative of the 1-form ${\omega}_i{\rm d} x^i$.
   The space-time is {\em static} when the form ${\omega}_i{\rm d} x^i$ is
exact. Furthermore in a static space-time the 1-form $G$ is real.

 The Einstein equations are
\begin{equation}  \label{eq:feq}
  E_{\mu\nu}{\stackrel{\rm def}{=}} R^{(4)}_{\mu\nu}+kT^{*}_{\mu\nu}=0 
  \qquad (\mu,\nu,... = 0,1,2,3) ,
\end{equation}
where
$T^{*}_{\mu\nu} {\stackrel{\rm def}{=}} T_{\mu\nu} - 
\slantfrac{1}{2}g_{\mu\nu}T$.
With these, the Eqs. (\ref{eq:fieldeqs}) take the
form\footnote{Vectors in the tangent 3-space are set in boldface.
A dot notation $(.)$ is used for
a scalar product with respect to the 3-metric.}
\begin{mathletters}\label{eq:3eq}
  \begin{eqnarray}
    {G^{ i}}_{;i} & = &({\bf{ G}.{ G}}) - ({\bf{ G}.\bar{ G}}) +
    kr^{-2}T^*_{oo} , \label{eq:divT} \\
    G_{ i;j}-G_{ j;i} & = & \bar G_{ i}G_{ j} - G_{ i}\bar
    G_{ j} + ikr^{-2}{\epsilon}_{ ijk}T^{* k}_o ,  \label{eq:commG} \\
    R_{ ij} & = & -G_{ i}\bar G_{ j} - \bar G_{ i}G_{ j} -
    kr^{-2}\left(T^*_{ ij}-g_{ ij}T^{*}_{oo}\right) , \label{eq:Rij}
  \end{eqnarray}
\end{mathletters}
 
  From the Bianchi identities
$(R^i_j- \slantfrac{1}{2}\delta^i_jR)_{;i}=0$,
in the 3-space, we then get
\begin{equation}\label{eq:genBian}
  \left[\frac1{r^2}\left(T^{*i}_j -
    \slantfrac{1}{2}\delta^i_jT^{*k}_k\right)\right]_{;i}
  +\frac1{2r^2}T^*_{oo;j}
  +i\frac1{r^2}{\epsilon_{ijk}}(G^i-\bar G^i)T^{*k}_o=0\ .
\end{equation}
Eqs. (\ref{eq:3eq}) and (\ref{eq:genBian}) are valid
for an arbitrary space-time admitting the Killing
vector $K$.
 
 An important example to be investigated in this paper is the
{\em perfect fluid} with the energy-momentum tensor
\begin{equation}
  T_{\mu\nu}=(\mu+p)u_{\mu}u_{\nu}-pg_{\mu\nu}\ .
\end{equation}
where $p$ is the pressure and $\mu$ the matter density.
The 4-velocity $u^{\mu}$ is normalized by $u^{\mu}u_{\mu}=1$.
This will be written as
\begin{equation} \label{eq:uu}
  u_{o}^2-u_k u^k=r \ .
\end{equation}
The energy-momentum tensor of the stationary perfect fluid becomes
\begin{mathletters}\label{eq:T}
  \begin{eqnarray}
    T^{*}_{oo}&=&(\mu+p)u_{o}^2-\frac{1}{2} (\mu-p)r , \\
    T^{*i}_{o}&=&(\mu+p)u_{o}u^i , \\
    T^{*ij}&=&(\mu+p)u^{i}u^j+\frac{1}{2} g^{ij} (\mu-p)r \ .
 \end{eqnarray}
\end{mathletters}

\section{Axisymmetry}

  A stationary axisymmetric space-time is defined by
the following properties:
\begin{enumerate}
  \item[(i)] There is a space-like Killing vector
    $\partial/\partial{\varphi}$ and a time-like Killing vector
    $K=\partial/\partial t$ along which all quantities are Lie
    transported.
  \item[(ii)] The metric has the axisymmetric form
    \begin{equation}
      {\rm d} s^2=r({\rm d} t+{\omega}
      {\rm d}{\varphi})^2-r^{-1}\{e^{2\lambda}[({\rm d} x^1)^2 + 
      ({\rm d} x^2)^2]+\rho^2
      {\rm d}{\varphi}^2\} ,
    \end{equation}
    and the metric functions $r,\ {\omega},\ {\lambda}$, and $\rho$
    depend on the coordinates $(x^A){\stackrel{\rm def}{=}}(x^1,x^2)$.
\end{enumerate}
We label the remaining coordinates as $t=x^0$ and ${\varphi}=x^3$.

\subsection{Matter tensor}

  The condition for the 3-space to be conformally flat is that the
Cotton-York tensor \cite{bible}
\begin{equation} \label{eq:york}
  Y_i^{\ell}{\stackrel{\rm def}{=}}{\epsilon^{jk\ell}}\left(R_{i[j;k]} - 
  \slantfrac{1}{4}g_{i[j}R_{;k]}\right) , 
\end{equation}
vanishes. The Cotton-York tensor is symmetric, trace- and
divergence-free:
\begin{eqnarray}
  Y_{[ij]}=0, & \ Y_{i}^i=0, & \ {{Y_i}^{j}}_{;j}=0 \ .
\end{eqnarray}
 
 In addition, we introduce the {\em complex} tensor
\begin{equation} \label{eq:simon}
  S_i^{\ell}={\epsilon^{jk\ell}}\left\{ \frac{}{}
            2g_{ij}g^{rs}G_{[k;|r|}G_{s]}-2G_{k;i}G_{j}
           -ikr^{-2}\epsilon_{jk}^{\ \ r}G_{(i}T^*_{or)}
                        \right\} ,
\end{equation}
which is similarly symmetric and trace-free
both in vacuum and in the presence of matter:
\begin{eqnarray}
  S_{[ij]}=0, & \ S_{i}^i=0\ .
\end{eqnarray}
The tracefree property follows from the fact that $G_i$ and
$R^{(4)i}_o$ are orthogonal.
In (\ref{eq:simon}), the constant $k$ is arbitrary, but later on
$k$ will be identified with the gravitational constant.
The tensor $S_i^{\ell}$ is real in a static space-time. In vacuo,
$S_i^{\ell}$ equals the Simon tensor \cite{Zol2,Simon}. Examples
of space-times with a vanishing $S_i^{\ell}$ are the (vacuum) Kerr
metric, the interior Schwarzschild, Kramer and Wahlquist
space-times \cite{bible,Kramer}.
 
  In all known examples of static metrics with a vanishing Simon
tensor, the Simon tensor equals the Cotton-York tensor. However,
as can be easily seen by inspecting the form of the Cotton-York
tensor (\ref{eq:york}), this is the case when the energy-momentum
terms separately vanish.
 
  We construct another symmetric and trace-free tensor for
stationary space-times with matter. We adopt the following
procedure:
\begin{enumerate}
  \item First we rewrite the Cotton-York tensor of a static
    space-time by use of the field equations (\ref{eq:feq}) and the
    decomposition (\ref{eq:ricdec}) of the Ricci tensor in terms of
    the real 1-form $G$ and matter variables.
  \item We then complexify this tensor for a nonstatic
    space-time by allowing $G$ to be complex and separate the
    tensor $S_{ik}$.
  \item By adding appropriate
    matter terms we ensure that the complexified tensor is symmetric
    and trace-free.
\end{enumerate}
 
 From step 1, we get the Cotton-York tensor in the static case,
\begin{equation}
  Y_i^{\ell}={\epsilon^{jk\ell}}\left\{
            2g_{ij}g^{rs}G_{[k;|r|}G_{s]}-2G_{k;i}G_{j}
            + k(r^{-2}\Delta_{ij})_{;k}
            -kr^{-2} g_{ij} G_k T^{*}_{oo} \right\} ,
\end{equation}
with
\begin{equation}
  \Delta_{ij} {\stackrel{\rm def}{=}} T^{*}_{ij}
            -\slantfrac{1}{4}g_{ij}T^{*}_{oo}
            -\slantfrac{1}{4} g_{ij}{T^{*}_r}^r \ .
\end{equation}
After having added compensating terms to ensure that the tensor is
trace-free and symmetric, there still remains a freedom of adding
a symmetric and trace-free tensor which vanishes in the static
limit. This latter tensor will be determined in Sec. \ref{sec:fluid}.
The matter tensor then reads
\begin{eqnarray}
  C_i^{\ell} &=& {\epsilon^{jk\ell}}\left\{
             - (r^{-2}\Delta_{ij})_{;k}
             +  r^{-2} g_{ij}(G_k+\bar G_k)T^{*}_{oo}/2  
             \right. \nonumber \\ 
            &&\left. + ir^{-2} {{\epsilon}_{ik}}^r (G_j- \bar G_j)
             T^{*}_{or}
             +4ir^{-2} {{\epsilon}_{ik}}^r (G_{(j}- \bar G_{(j})
             T^{*}_{or)} \right\} \ . \label{eq:C}
\end{eqnarray}

\subsection{A complex basis}\label{sec:ComplexBasis}
 
  From property (i), ${\bf L} =\partial/\partial{\varphi}$ is a Killing
vector in the 3-space and
$({\bf G} . {\bf L})=0$.
 In the generic case, the triad
 $({\bf G} ,\bar {\bf G} , {\bf L} )$ represents a proper frame basis
in the tangent 3-space.
However, we shall use another complex basis\footnote{Component
  indices in the basis (\ref{eq:basis}) are set in boldface.}
\begin{equation} \label{eq:basis}
 ({\bf e}_{\bf 1},{\bf e}_{\bf 2},{\bf L} ).
\end{equation}
The vectors ${\bf e}_{\bf 1}$ and ${\bf e}_{\bf 2}$ form a complex
conjugate pair, ${\bf e}_{\bf 2}=\bar {\bf e}_{\bf 1}$ and they are
defined by a rotation in the $({\bf G},\bar {\bf G})$ subspace
\begin{mathletters}
  \begin{eqnarray}
         {\bf G}&=&\frac{1}{2r}({\alpha} {\bf e}_{\bf 1} +
         {\beta}{\bf e}_{\bf 2}) , \\
    \bar {\bf G}&=&\frac{1}{2r}({\beta} {\bf e}_{\bf 1} +
         {\gamma}{\bf e}_{\bf 2}) , 
  \end{eqnarray}
\end{mathletters}
with
\begin{equation}
  {\alpha}=4r^2({\bf G}.{\bf G}),\qquad 
  {\beta}=4r^2({\bf G}.\bar {\bf G}),
  \qquad {\gamma}=\bar{\alpha}\ .
\end{equation}
  The metric in this basis reads
\begin{equation} \label{eq:g^ik}
  [{g^{\bf ik}}]= \left[ \matrix{
        {\alpha}   &   {\beta}  &   0 \cr
        {\beta}   &   {\gamma}  &   0 \cr
         0    &    0   &   \rho^{-2} \cr } \right] ,
\end{equation}
where $\rho^2=({\bf L}.{\bf L})$
is the squared length of the Killing vector. Note that the
basis (\ref{eq:basis}) is neither holonomic, nor is it rigid.
 
  The {\em structure functions} ${c^{\bf i}}_{\bf jk}$ are defined
by the commutator
\begin{equation} \label{eq:comm}
  [{\bf e_j},{\bf e_k}] = {c^{\bf i}}_{\bf jk}{\bf e_i} \ .
\end{equation}
In the absence of matter, the basis (\ref{eq:basis}) is natural and
the structure functions vanish.

For the directional derivatives we use the alternative notation with a
comma in the subscript. For example, we have from (\ref{eq:G}):
\begin{equation}
  {\bf e}_{\bf1} r \equiv r_{\bf,1} = \slantfrac{1}{2},\qquad
  {\bf e}_{\bf2} r \equiv r_{\bf,2} = \slantfrac{1}{2} \ .
\end{equation}
 
The dual basis reads $({\theta}^{\bf 1}, {\theta}^{\bf 2},
{\theta}^{\bf 3}) = (2rG,\ 2r\bar G,\ \rho^{-2}L)$. In this basis, 
$G$ has the components
\begin{mathletters}
  \begin{eqnarray}
   &(G_{\bf1},G_{\bf2},G_{\bf3}) = \left(\frac1{2r},0,0\right) , \qquad
    (\bar G_{\bf1},\bar G_{\bf2},\bar G_{\bf3})
                    =\left(0,\frac{1}{2r},0\right) , \label{eq:G_i}\\
   &(G^{\bf1},G^{\bf2},G^{\bf3}) = 
    \left(\frac{{\alpha}}{2r} , \frac{{\beta}}{2r}
     , 0\right) , \qquad 
   (\bar G^{\bf1},\bar G^{\bf2},\bar G^{\bf3})
     =\left(\frac{{\beta}}{2r},\frac{{\gamma}}{2r},0\right)&\ .
   \label{eq:G^i}
  \end{eqnarray}
\end{mathletters}
Because we use a frame of nonconstant normalization,
we need the structure functions to compute the connection
coefficients. A short calculation using
\begin{equation}
  {\rm d}{\theta}^{\bf i} =
  -\slantfrac{1}{2}{c^{\bf i}}_{\bf jk}{\theta}^{\bf
  j}{\wedge}{\theta}^{\bf k} ,
\end{equation}
and (\ref{eq:commG}) in natural
coordinates gives the structure functions
\begin{mathletters}
  \begin{eqnarray}
    {c^{\bf 1}}_{\bf j\bf k}&=&-{c^{\bf 2}}_{\bf jk} =
    2ikr^{-1}{\epsilon}_{\bf j kl}T^{*\bf l}_{o} , \\
    {c^{\bf 3}}_{\bf j\bf k} &=&0\ .
  \end{eqnarray}
\end{mathletters}
There is only one independent structure function which we denote
${c^{\bf1}}_{\bf12}{\stackrel{\rm def}{=}}{\varepsilon}$. We have
\begin{equation}
  {\varepsilon} = 2i\frac kr\frac{\rho}{\sqrt{D}}T^{*{\varphi}}_o , 
\end{equation}
where $D={\alpha}{\gamma}-{\beta}^2$.
We obtain the detailed form of Eq. (\ref{eq:divT}) by computing
the contracted connection coefficients
$\Gamma^{\bf i}_{\bf ki} = {\bf e}_{\bf k}(\ln
\rho/\sqrt D)+{c^{\bf i}}_{\bf ik}$.
With the components (\ref{eq:G^i}) of $\bf G$, the field equation
(\ref{eq:divT}) takes the form
\begin{equation} \label{eq:divergence}
  {\alpha}_{,\bf1}+{\beta}_{,\bf2}-\frac{{\alpha}}{r} +
  ({\alpha} {\bf e}_{\bf1}
  + {\beta} {\bf e}_{\bf2})\ln\frac{\rho}{\sqrt D} - 
  2{\frac kr}T^*_{oo} + ({\alpha}+{\beta}){\varepsilon}=0 . 
\end{equation}
Eqs. (\ref{eq:commG}) are now absorbed in the structure
equations (\ref{eq:comm}). Eqs. (\ref{eq:Rij}) are given in
Appendix A.
 
\subsection{Fields with
$S_{\lowercase{i}}^{\ell}=0$}\label{sec:ZeroSimon}
 
 The two nonvanishing components of the tensor $S_i^{\ell}$ may be
written
\begin{equation} \label{eq:U}
  U_i{\stackrel{\rm def}{=}} -4{\epsilon_{ijk}} L^j S^{k\ell}L_{\ell} ,
\end{equation}
where $({\bf U.L})=0$.
  The contents of the condition $U_{\bf2}=0$ for vacuum is
${\alpha}_{,\bf2}=0$ \cite{Zol2}. We may preserve the vacuum form of
$U_{\bf2}$ in the presence of matter by identifying the
constant $k$ in (\ref{eq:simon}) with the gravitational constant.
 
   Substituting the definition (\ref{eq:simon}) of the tensor
$S^{k\ell}$ in Eq. (\ref{eq:U}) and
using the field equations (\ref{eq:divT})
and (\ref{eq:commG}), we get
\begin{eqnarray}
  U_i&=&-4\{-(\rho^2)_{,a}G^aG_i+\rho^2[({\bf G.G})+kr^{-2}T^*_{oo}]G_i
                                                   \nonumber\\&&
    -\rho^2[(1/2) ({\bf G.G})_{,i}+({\bf G.G})\bar G_i]
    +ikr^{-2}{\epsilon_{ijk}} G^k T_o^{*\ell}(L_{\ell}L^j - 
    \delta_{\ell}^j\rho^2)\} \ .
\end{eqnarray}
The last term contains a projector in the orthogonal complement of the
Killing vector $L$, thus it vanishes. Taking the triad components, we
obtain
\begin{mathletters}\label{eq:nonhoU}
  \begin{eqnarray}
    U_{\bf1}&=&2\frac{\rho^2}{r^2}\left[
    ({\alpha}{\bf e}_{\bf1}+{\beta}{\bf e}_{\bf2})\ln{\rho}
    -\frac{{\alpha}}{2r}+\frac{{\alpha}_{,\bf1}}{4}-\frac{k}{r}
    T^*_{oo}\right], \label{eq:u1} \\
    U_{\bf2}&=&
    \frac{\rho^2}{2r^2}{\alpha}_{,\bf2} . \label{eq:u2}
  \end{eqnarray}
\end{mathletters}
Putting the tensor $S_{ik}$ to zero gives, through Eq. (\ref{eq:u1}),
\begin{equation}
  ({\alpha}{\bf e}_{\bf1}+{\beta}{\bf e}_{\bf2})\ln{\rho}
  -\frac{{\alpha}}{2r}+\frac{{\alpha}_{,\bf1}}{4}-\frac{k}{r}
  T^*_{oo} = 0 .   \label{eq:rh1}
\end{equation}
We now subtract this equation from (\ref{eq:divergence}) as to
eliminate the derivatives of $\rho$. The equation obtained is then
multiplied with ${\beta}$ and its complex conjugate with $-{\alpha}$.
Adding the results gives
\begin{mathletters}\label{eq:beder}
  \begin{eqnarray}
    4{\alpha}{\beta}_{,\bf1} - 3{\beta}{\alpha}_{,\bf1} + 
    {\alpha}{\gamma}_{,\bf2}
    + 2\frac{{\alpha}}{r}({\beta}-{\gamma}) + 4\frac{k}{r}({\beta} - 
    {\alpha})T^*_{oo}
                            + 4D{\varepsilon}&=&0 , \label{eq:419}  \\
    4{\gamma}{\beta}_{,\bf2} - 3{\beta}{\gamma}_{,\bf2} + 
    {\gamma}{\alpha}_{,\bf1}
    + 2\frac{{\gamma}}{r}({\beta}-{\alpha}) + 4\frac{k}{r}({\beta} - 
    {\gamma})T^*_{oo}
                            + 4D{\varepsilon}&=&0 \label{eq:cc419} \ .
 \end{eqnarray}
\end{mathletters}
 
We can also obtain an equation for $\rho_{,\bf1}$, by simply multiplying
(\ref{eq:rh1}) by ${\gamma}$ and its complex conjugate by $-{\beta}$
and then taking the sum:
\begin{equation} \label{eq:drho}
  D(\ln\rho)_{,\bf1}+\slantfrac{1}{4}({\gamma}{\alpha}_{,\bf1} - 
  {\beta}{\gamma}_{,\bf2}) +
  \frac{{\gamma}}{2r}({\beta} - {\alpha}) -\frac{k}{r}({\gamma} - 
  {\beta})T^*_{oo} = 0 .
\end{equation}
 
Combining Eqs.
(\ref{eq:drho}) and (\ref{eq:419}) as to eliminate the 
${\bf e}_{\bf2}$
derivatives we obtain
\begin{equation}
  {\alpha}(\ln c)_{,\bf1}=F \ ,
\end{equation}
where
\begin{equation} \label{eq:F}
  F{\stackrel{\rm def}{=}}(k/r)T^*_{oo}-{\beta}{\varepsilon} ,
\end{equation}
and the real function $c{\stackrel{\rm def}{=}} 
D^{-1/2}r^{-1}({\alpha}{\gamma})^{3/4}\rho$ is known
to be constant in the vacuum case \cite{Zol2}.
Notice that $F$ does not vanish for the
interior Schwarzschild space-time.
Applying the commutator of the derivatives on $c$, we have
\begin{equation} \label{eq:Too2}
  {\alpha} F_{,\bf1}-{\gamma} F_{,\bf2} = 
  {\varepsilon}({\gamma}-{\alpha})F \ .
\end{equation}
 
Upon multiplying (\ref{eq:cc419}) by $-1/4$ and
adding it
to (\ref{eq:drho}) we eliminate ${\alpha}_{,\bf1}$.
We then have
\begin{mathletters}\label{eq:rhoder}
  \begin{eqnarray}
    D(\ln\rho)_{,\bf1} + \slantfrac{1}{2}{\beta}{\gamma}_{,\bf2} - 
    {\gamma}{\beta}_{,\bf2}
    - D{\varepsilon} & = & 0 \ , \label{eq:rho1}\\
    D(\ln\rho)_{,\bf2} + \slantfrac{1}{2}{\beta}{\alpha}_{,\bf1} - 
   {\alpha}{\beta}_{,\bf1}
    - D{\varepsilon} & = & 0 \ . \label{eq:rho2}
  \end{eqnarray}
\end{mathletters}
 
  By applying the commutator (\ref{eq:comm}) to the field variables,
we obtain additional conditions. We eliminate the function $\rho$
from what follows by use of Eqs. (\ref{eq:rhoder}).
For instance, applying (\ref{eq:comm}) on the function 
${\beta}$, we obtain
a second order equation which we use in turn for eliminating the
derivative ${\gamma}_{,\bf22}$. Similarly, from Eq. (\ref{eq:Too2}),
we may eliminate the derivatives $T^*_{oo,\bf2}$. At this juncture, we
find that the integrability conditions of the function $\rho$ are
satisfied identically.
Following this, we use Eqs. (\ref{eq:beder}) for the elimination of
the derivatives ${\beta}_{,\bf2}$ and ${\gamma}_{,\bf2}$.
As a result of this, all ${\bf e}_{\bf2}$
derivatives of the variables may be expressed in terms of
${\bf{\varepsilon}_{,2}}$,
the ${\bf e}_{\bf1}$ derivatives and the variables themselves. We
may proceed to get rid of the remaining ${\bf e}_{\bf2}$ derivative
by use of the field equations (\ref{eq:Rij}). This will enable us to
obtain a system of `ordinary' differential equations containing only
${\bf e}_{\bf1}$ derivatives.
 
  The above procedure is valid for an arbitrary medium.
In the following, we develop this approach further for a perfect fluid.

\section{Perfect Fluid}\label{sec:fluid}
 
   In accordance with property (i), the pressure $p$, matter density
$\mu$ and four-velocity $u^{\mu}$ of an axisymmetric stationary perfect
fluid depend on the two coordinates $x^A$.
 From the field equation (\ref{eq:commG}) it follows that the
three-velocity has the form $u=u_{{\varphi}}{\rm d}{\varphi}$, thus it is
expansion-free,
$u^i\!_{;i}=(u^{{\varphi}})_{,{\varphi}}=0$.
The normalization condition (\ref{eq:uu}) for the 4-velocity of
the fluid becomes
\begin{equation} \label{eq:norm}
  u_{o}^2-\rho^{-2}u_{{\varphi}}^2=r \ .
\end{equation}
 
 The following relations hold for the components (\ref{eq:T})
of the energy-momentum tensor:
\begin{equation}\label{eq:T33}
  T^{*}_{{\varphi}{\varphi}}-\rho^2(T^{*}_{oo}-2rp)=0 ,
\end{equation}
and
\begin{equation}\label{eq:alg}
  r^2D{\varepsilon}^2+k^2[2T^{*}_{oo}+(\mu-p)r][2T^{*}_{oo}
  -(\mu+3p)r]=0\ .
\end{equation}
 
     The Bianchi identities (\ref{eq:genBian}) take the form
\begin{equation}\label{eq:Bianchi}
{\rm d} p=-\frac{\mu+p}{2r^2\rho^3}
   [2ru_{{\varphi}}(u_or\rho{\rm d}{\omega}-u_{{\varphi}}{\rm d}\rho)
   +(r\rho^2+2u_{{\varphi}}^2)\rho {\rm d} r] \ .
\end{equation}
Because of the structure of the field equations, there are only
two nontrivial components.\footnote{The contraction of the
conservation law with the Killing vector $K$ evaporates for axisymmetry:
$K_{\mu}{T^{\mu\nu}}_{\vert\nu}=(K_{\mu}T^{\mu\nu})_{\vert\nu}=0$.}
The integrability condition of ${p}$ yields
\begin{equation}\label{eq:intp}
  {\rm d} [u_o(\mu+p)]{\wedge}{\rm d}[{\omega}-u_o/(ru^{{\varphi}})]
  +(u^{{\varphi}})^{-2} {\rm d} u^{{\varphi}}{\wedge}{\rm d}\mu=0 \ .
\end{equation}
This generalizes {\it von Zeipel's theorem} in general relativity
\cite{Thorne}, according to which $u_o(\mu+p)=const.$ for a static
fluid. For a rigidly rotating fluid (${\rm d} u^{{\varphi}}=0$), we have
${\omega}=u_o/(ru^{{\varphi}})+const$.
Some related Hamiltonian structures have been discussed by
Stephani and Grosso \cite{Stephani}.
 
  The matter tensor $C_i^{\ell}$ [Eq. (\ref{eq:C})] is fully determined
by requiring that the components of the tensor are integrable.
The condition that the tensor $C_i^{\ell}$ vanishes may be integrated to
yield
\begin{equation}\label{eq:intC}
  (\mu+p)r^{-2}\rho^{-1}u_{{\varphi}}^2=const.
\end{equation}
 
  The fields with a vanishing $S_i^{\ell}$ tensor  [Eq. (\ref{eq:simon})]
merit further investigation.

\subsection{Perfect fluids with $S_i^{\ell}=0$}
 
The equation ${\alpha} R_{\bf11}-{\gamma} R_{\bf22}=0$ for 
the Ricci tensor components of the {\em perfect fluid} yields
\begin{equation} \label{eq:dep}
  {\gamma}{\varepsilon}_{,\bf2}-{\alpha}{\varepsilon}_{,\bf1} = 
  \left[\left({\varepsilon}-\frac1{2r}\right)({\alpha}-{\gamma})
                             +\frac34({\alpha}_{\bf1} - 
  {\gamma}_{\bf2})\right]{\varepsilon} ,
\end{equation}
which we use for eliminating ${\varepsilon}_{,\bf2}$. 
The commutator on $c$ thus simplifies to the form:
\begin{equation}  \label{eq:T2}
  k({\gamma} T^*_{oo,\bf2}-{\alpha} T^*_{oo,\bf1})=
  \frac{k}{2r}({\gamma} - {\alpha})T^*_{oo}
          +\frac14{\varepsilon} r({\alpha}{\gamma}_{,\bf2} - 
  {\gamma}{\alpha}_{,\bf1})\ .
\end{equation}
 
   The Bianchi identities (\ref{eq:Bianchi}) have the
only independent complex triad component
\begin{equation}\label{eq:p1}
  p_{\bf,1}=\frac{1}{2r}\left\{p+\frac{1}{2k}{\varepsilon}
    ({\beta}-{\gamma})-\frac{1}{r}T^*_{oo}
   +[2T^*_{oo}-(\mu+3p)r]\frac{\rho_{\bf,1}}{\rho}\right\} \ .
\end{equation}
 
  Carrying out the above substitutions in the field equation 
(\ref{eq:r33}), we get
\begin{eqnarray}  
  {\alpha}{\alpha}_{,\bf11}-\frac{3}{4}({\alpha}_{,\bf1})^2 & = & 
  4{\alpha}(2{\varepsilon}{\beta}_{,\bf1} +
  {\beta}{\varepsilon}_{,\bf1} + F_{,\bf1}) - 2(3{\beta}{\varepsilon} + 
  2F){\alpha}_{,\bf1}
  \nonumber \\
  &&+ 4F(2{\beta}{\varepsilon} + F) + 
  4\frac{{\alpha}}{r}({\beta}{\varepsilon}
  + r{\gamma}{\varepsilon}^2 + F - kp) \ .  \label{eq:E00}
\end{eqnarray}

 We next eliminate terms quadratic in the gravitational
constant, by use of Eq. (\ref{eq:r33}), from  Eqs. (\ref{eq:r11}) and
(\ref{eq:r12}). The latter equation  % $(E_{12})$ 
then takes the form
\begin{equation}\label{eq:F1}
  F_{,\bf1}+{\varepsilon} F={\gamma}{\varepsilon}^2+
  \frac{{\gamma}}{2Dr}[k(\mu+3p){\beta}-({\alpha}{\gamma}
  +{\beta}^2){\varepsilon}-2F{\beta}]\ .
\end{equation}
Eq. (\ref{eq:Too2}) is just the imaginary part of this.
We now get rid of $T^*_{oo}$ using the definition (\ref{eq:F})
of $F$ and cancel the terms linear in $k$, using (\ref{eq:F1}) in
% $(E_{11})$
(\ref{eq:r11}). Simplifying by an overall positive factor $\mu+3p$, we
have
\begin{equation}\label{eq:E11}
  {\alpha} F_{,\bf1}+({\alpha}-{\beta}){\varepsilon} F
   ={\alpha}{\varepsilon}({\gamma}-{\beta})\left({\varepsilon} - 
  \frac1{2r}\right)
                -\frac34{\beta}{\varepsilon}{\alpha}_{,\bf1}
                -{\alpha}{\beta}{\varepsilon}_{,\bf1}\ .
\end{equation}
 The field equations (\ref{eq:alg}), (\ref{eq:p1}),
(\ref{eq:E00}), (\ref{eq:F1}) and (\ref{eq:E11}) describe a
rotating perfect fluid with the tensor $S^i_k$ vanishing. They are
a set of partial differential equations for the complex
variables ${\alpha}$ and ${\gamma}=\bar{\alpha}$ and the real variables 
${\beta},\ p,\ {\varepsilon},\ F$ (substituting for $T^*_{oo}$)
 and $r$ with 
$r_{,\bf1}=1/2$.
These field equations contain the only derivative operator
${\bf e}_{\bf1}$.
The right hand side of Eq. (\ref{eq:E00}) vanishes in vacuum, and we
obtain Simon's \cite{Simon} {\it eikonal equation} in that limit.
The remaining equations of the system contain only matter variables,
thus they become identities in vacuum.
 
In the next section, we shall
investigate incompressible perfect fluids with an arbitrary
$S_i^{\ell}$ tensor. In Sec. \ref{sec:ConstantAl}, we give an illustrative
example of solving the field equations for the simple case when the
function ${\alpha}$ is constant.
 
\subsection{Incompressible fluid}
 
We eliminate $\rho$ from the Bianchi identities (\ref{eq:Bianchi}),
using the normalization condition (\ref{eq:norm}), to
obtain
\begin{equation} \label{eq:IncomprBian}
  {\rm d} p=(\mu+p)\left[ u_o{\rm d}\left(\frac{u_o}{r}\right) -
  \frac{u_o^2}{ru^{{\varphi}}}{\rm d} u^{{\varphi}} + 
  \frac{{\rm d} u^{{\varphi}}}{u^{{\varphi}}} -
   u^{{\varphi}}u_o{\rm d}{\omega} \right] \ .
\end{equation}
In the case of an {\em incompressible fluid} (${\rm d}\mu=0$), the
integrability condition
(\ref{eq:intp}) takes on a simpler form. We use Eq.
(\ref{eq:IncomprBian})
for substituting ${\rm d}(\mu+p)$. After expanding the terms, we get
\begin{equation}
  {\rm d}(u^{{\varphi}}u_o){\wedge}{\rm d}\left({\omega} - 
  \frac{u_o}{ru^{{\varphi}}}\right)=0\ .
\end{equation}
Hence ${\omega}$ has the functional form \cite{Abram}
\begin{equation}
  {\omega} = \frac{u_o}{ru^{{\varphi}}} + \Psi(\chi) ,
\end{equation}
where $\Psi$ is some functional of $\chi {\stackrel{\rm def}{=}} 
 u_ou^{{\varphi}}$. Inserting
this in (\ref{eq:IncomprBian}) gives us
\begin{equation}
  {\rm d} \ln[(\mu + p)(u^{{\varphi}})^{-1}] + \chi{\rm d}\Psi=0 \ .
\end{equation}
Redefining the functional by
\begin{equation}
  \Phi(\chi) {\stackrel{\rm def}{=}}
  \int{\chi}\frac{\delta\Psi}{\delta\chi}{\rm d}\chi ,
\end{equation}
enables us to integrate the pressure in an explicit form:
\begin{equation}
  \mu+p= u^{{\varphi}}e^{-\Phi(\chi)} \ .
\end{equation}
Thus for incompressible fluids we may express the pressure in terms of
the 4-velocity components.

\subsubsection{Incompressible fluid with $S_i^{\ell} = 0$}

   We now introduce the nonholonomic basis (\ref{eq:basis})
for the subsequent discussion of the incompressible fluid. Our aim
is to prove that the derivative ${\alpha}_{,\bf1}$ is real.
 
   We take the ${\bf e}_{\bf2}$ derivative of the Einstein
equation  
% ($E_{00}$)
(\ref{eq:E00}), and eliminate ${\gamma}_{,\bf2}$, ${\beta}_{,\bf2}$,
$p_{,\bf2}$, ${\varepsilon}_{,\bf2}$,
and $F_{,\bf2}$ by (\ref{eq:419}), (\ref{eq:cc419}),
the complex conjugate of (\ref{eq:p1}), (\ref{eq:dep}),
and (\ref{eq:Too2}), respectively.
Next we go on deriving further relations, from which we
systematically
eliminate $p_{,\bf1}$, ${\varepsilon}_{,\bf1}$, and $F_{,\bf1}$ by the
respective use of Eqs. (\ref{eq:p1}),  
(\ref{eq:E11}) and (\ref{eq:F1}).  % ($E_{12}$)
We obtain two first-order equations by taking
first the ${\bf e}_{\bf1}$ derivative of (\ref{eq:alg}), and then acting
with the commutator on (\ref{eq:alg}). Acting with the commutator on
${\beta}$, we get a second-order equation. The resultant of this and 
the ${\bf e}_{\bf2}$ derivative of Eq. (\ref{eq:E00})
% ($E_{00}$)
 with respect to ${\beta}_{,\bf12}$ may be
simplified by a common factor $D$. Taking again the resultant of
Eq. (\ref{eq:E00}) with respect to ${\alpha}_{,\bf12}$, we get 
the first-order equation
\begin{equation}
  4{\alpha}{\beta}_{,\bf1} - 3{\beta}{\alpha}_{,\bf1} + 
  {\alpha}{\alpha}_{,\bf1}
  + 2\frac{{\alpha}}{r}({\beta}-{\gamma}) + 4\frac{k}{r}({\beta} - 
  {\alpha})T^*_{oo} + 4D{\varepsilon} = 0\ .
\end{equation}
Comparing with (\ref{eq:419}), we obtain the reality condition
\[
 {\alpha}_{,\bf1}={\gamma}_{,\bf2} .
\]

\subsubsection{Solution for constant ${\alpha}$}\label{sec:ConstantAl}

   As an illustrative example how to solve our field equations with
$S_i^{\ell} = 0$,
we now obtain an incompressible solution, starting with the simple
assumption that the function ${\alpha}$ and its complex conjugate 
${\gamma}$ are constants.
 
   We express the first derivatives of the function ${\beta}$ from Eqs.
(\ref{eq:beder}). Acting with the commutator
(\ref{eq:comm}) on the condition (\ref{eq:alg}), we get
\begin{equation} \label{eq:a12}
  {\varepsilon}({\alpha}{\gamma}+{\beta}^2-4{\beta}^2{\varepsilon} r - 
  4{\beta} Fr)
  +({\beta}-2rF-2{\beta}{\varepsilon} r)(2F-k\mu-3kp)=0 \ .
\end{equation}
The integrability condition of ${\beta}$ yields
the algebraic relation
\begin{equation}\label{eq:b12}
  {\alpha}{\gamma}(2r{\varepsilon}-1)+2rF{\beta}=0 \ .
\end{equation}
We eliminate the function ${\beta}$ from here by use of
({\ref{eq:a12}). The resulting algebraic
equation splits in two factors. One of the factors can be shown
nonvanishing. The other yields
\begin{equation}
  -{\alpha}{\gamma}{\varepsilon}(4r{\varepsilon}-1)+rF(4F-2k\mu-6kp)=0 \ .
\end{equation}
Similarly eliminating ${\beta}$ between Eq. (\ref{eq:b12}) and
(\ref{eq:Too2}), we get a large polynomial expression vanishing.
The expression factorizes again. Dropping the nonvanishing factors,
we have
\begin{equation}
  {\alpha}{\gamma}{\varepsilon}(6{\varepsilon}^2r^2-2{\varepsilon} r-1) - 
  rF(6rF{\varepsilon}+4F+4k\mu)=0\ .
\end{equation}
 From this we obtain
\begin{equation}
  {\beta}=(2/3) k\mu(1/{\varepsilon}-2r) \ .
\end{equation}
Completing the procedure, we get the explicit form of the field
variables as functionals of $r$:
\begin{eqnarray*} \label{consal}
  {\beta}  = -[9{\alpha}{\gamma}+(4k\mu r)^2]/(24k\mu r)    , \qquad
  {\varepsilon} = -(4k\mu)^2 r/\Lambda , \qquad
    F  = -12{\alpha}{\gamma} k\mu/\Lambda , \qquad
    p  = -\mu                                 ,
\end{eqnarray*}
where $\Lambda(r) {\stackrel{\rm def}{=}} 9{\alpha}{\gamma}-(4k\mu r)^2$. 
The last equation shows that
the pressure is a negative constant, thus 
this particular solution does not describe a physically acceptable
fluid interior. It follows further from $\mu+p=0$ with
(\ref{eq:intC}) that the matter tensor $C_i^{\ell}$ vanishes.

\subsection{Integrability conditions}\label{sec:IntCond}

 In this section we investigate the integrability conditions for
incompressible fluid fields with a vanishing $S_i^{\ell}$ tensor
[Eq. (\ref{eq:simon})]. 
 From the integrability conditions of the Simon tensor and the
Bianchi identities, we obtain four real algebraic equations.
 Our algebraic conditions are real and contain the complex function
${\alpha}$ only through its the magnitude-square ${\alpha}{\gamma}$.
The equations are all quadratic in the fluid velocity
function ${\varepsilon}$, thus they have the form
\begin{equation} \label{eq:quad}
  Q_{{\alpha}} {\stackrel{\rm def}{=}} a_{{\alpha}}{\varepsilon}^2 + 
  b_{{\alpha}}{\varepsilon} + c_{{\alpha}}=0\ ,
\end{equation}
where ${\alpha} = 1, 2, 3, \text{ or } 4$.
The coefficients are polynomials, up to the 7$^{th}$ degree, in the
functions ${\beta}$, ${\alpha}{\gamma}$ and
\begin{equation}
  \psi {\stackrel{\rm def}{=}} \frac{k\mu r}{{\varepsilon} D}[2F + 
  2{\varepsilon}{\beta}-k(\mu+3p)]\ .
\end{equation}
For a rotating fluid, ${\varepsilon}\neq0$, it follows from the algebraic
constraint (\ref{eq:alg}) that the function $\psi$ must not
identically vanish.
We may express the pressure from the algebraic condition
(\ref{eq:alg}) as follows:
\begin{equation}
  p = -\frac{D{\varepsilon}\psi^2 + k^2\mu^2 r^2{\varepsilon} + 
  2k^2\mu^2r\psi}{2k^2\mu r\psi} \ . 
\end{equation}
The detailed form of the coefficients in (\ref{eq:quad}) is given in
Appendix C.
 
  We now bring the system of algebraic equations to a simpler form
by taking combinations of them.
We form the following system of vanishing polynomial expressions:
\begin{eqnarray*}
  {\rm  A } &{\stackrel{\rm def}{=}}& \frac12 Q_1 , \\
  {\rm  B } &{\stackrel{\rm def}{=}}& {\beta}\left[ \frac1{r\psi}\left(
          \frac{ Q_2}{1024m^2} - r^2 Q_4\right) -
          \psi Q_3 \right]  , \\
  {\rm  C } &{\stackrel{\rm def}{=}}&  Q_3 , \\
  {\rm  D } &{\stackrel{\rm def}{=}}& \frac1r\left\{ 9\left[rQ_
           4 + \frac{\psi}{r} Q_2 \right]
    - 2\left[ \frac1r\left( \frac{Q_2}{1024m^2} - r^2 Q_4
      \right) + \psi^2 Q_3 \right]\right\} , 
\end{eqnarray*}
where we have introduced the shorthand notation $m {\stackrel{\rm def}{=}}
 k\mu$.
 
We may lower the degree of these polynomials in $\psi$ by
rescaling the functions ${\alpha},{\beta}$ and ${\gamma}$:
${\beta}=b/\psi$, ${\alpha}{\gamma}=a^2/\psi^2$ where $a$ and $b$ are real
functions. With this, the Eqs.
 B  and  C  become linear and  A  is quadratic in $\psi$.
Eliminating next $\psi$ between  B  and  C , we obtain an
equation expressing the vanishing of a polynomial.
The polynomial may be split in two factors, one of which
however does not vanish. Thus we obtain the equation
  \begin{eqnarray}
    && 7 a^6 b r {\varepsilon} + 3 a^6 b + 8 a^6 m
    r^3 {\varepsilon}^2 - 13 a^4 b^3 r {\varepsilon} - 6 a^4 b^3 - 
    24 a^4 b^2 m r^3 {\varepsilon}^2 +
    4 a^4 b^2 m r \nonumber\\ &&
    + 16 a^4 b m^2 r^4 {\varepsilon}^2 - 3 a^4 b m^2 r^3 {\varepsilon}
    + 10 a^4 b m^2 r^2 - 16 a^4 m^3 r^5 {\varepsilon}^2 + 
    8 a^4 m^3 r^4 {\varepsilon}
    \nonumber\\ &&
    + 5 a^2 b^5 r {\varepsilon} + 3 a^2 b^5 + 
    24 a^2 b^4 m r^3 {\varepsilon}^2 - 4 a^2 b^4 m r
    - 32 a^2 b^3 m^2 r^4 {\varepsilon}^2 - 
    2 a^2 b^3 m^2 r^3 {\varepsilon} \nonumber \\ &&
    - 10 a^2 b^3 m^2 r^2 +
    32 a^2 b^2 m^3 r^5 {\varepsilon}^2 - 8a^2b^2m^3r^4{\varepsilon} - 
    4 a^2 b^2 m^3
    r^3 - 16 a^2 b m^4 r^6 {\varepsilon}^2 \nonumber\\ &&
    + 13 a^2 b m^4 r^5 {\varepsilon} - a^2 b m^4 r^4
    + 8 a^2 m^5 r^7 {\varepsilon}^2 - 8 a^2 m^5 r^6 {\varepsilon} +
    b^7 r {\varepsilon} - 8 b^6 m r^3
    {\varepsilon}^2 + 16 b^5 m^2 r^4 {\varepsilon}^2 \nonumber\\ &&
    + 5 b^5 m^2 r^3 {\varepsilon} - 16 b^4 m^3 r^5
    {\varepsilon}^2 + 16 b^3 m^4 r^6 {\varepsilon}^2 -
    13 b^3 m^4 r^5 {\varepsilon} - 8 b^2 m^5 r^7
    {\varepsilon}^2 + 7 b m^6 r^7 {\varepsilon}=0 , \label{p2}
  \end{eqnarray}
quadratic in ${\varepsilon}$.
We employ this relation to make the original four equations
linear in ${\varepsilon}$. We then find that some nonvanishing 
factors may be taken out. For example, both  B  and  C  may be divided
through $d=a^2-b^2$, and  C  is further divisible by
$f {\stackrel{\rm def}{=}} a^2-b^2-m^2r^2$. Having carried out these 
simplifications,
we eliminate the terms containing the product ${\varepsilon}\psi$ between
 B  and  C . We obtain a lengthy equation, which, however
can be factorized in the form
\begin{equation} \label{eq:BC}
  32qa^2b^2m^3r^4 f d (9 b-2 m r) (a+b-m r)^2 (a-b+m r)^2 = 0 , 
\end{equation}
where the factor $q$ is linear both in $\psi$ and
${\varepsilon}$:
  \begin{eqnarray}
    q &=& - 4 a^6 {\varepsilon} r - a^6 + 12 a^4 b^2 {\varepsilon} r + 
    3 a^4 b^2 + 12 a^4 {\varepsilon}
    m^2 r^3 - 16 a^4 m^2 \psi r^2 - 7 a^4 m^2 r^2 - 
   12 a^2 b^4 {\varepsilon} r
    \nonumber\\ &&
    - 3 a^2 b^4 - 24 a^2 b^2 {\varepsilon} m^2 r^3 
    + 32 a^2 b^2 m^2 \psi r^2 
    + 14 a^2
    b^2 m^2 r^2 + 24 a^2 b m^3 r^3 
    + 4 a^2 {\varepsilon} m^4 r^5 \nonumber\\ &&
    + 32 a^2 m^4\psi r^4
    + 37 a^2 m^4 r^4 + 4 b^6 {\varepsilon} r + b^6 + 
    12 b^4 {\varepsilon} m^2 r^3 -
    16 b^4 m^2 \psi r^2 - 7 b^4 m^2 r^2 - 24 b^3 m^3 r^3
    \nonumber\\ &&
    - 4 b^2 {\varepsilon} m^4 r^5
    - 32 b^2 m^4 \psi r^4 - 37 b^2 m^4 r^4 - 24 b m^5 r^5 
    - 12 {\varepsilon}
    m^6 r^7 - 16 m^6 \psi r^6 - 5 m^6 r^6 \ . \label{lin} 
  \end{eqnarray}
The Eq.  D=0 may be shown to be an algebraic consequence of the
other three equations.
 
   Eq. (\ref{eq:BC}) contains two factors of which the vanishing
is not physically unreasonable in the static limit. The
first alternative is that the factor $q=0$, and the other is that
$a-b+m r=0$. With the assumption that $q=0$, the solution of the
algebraic equations is a bothersome procedure yielding that the
junction surface $p=0$ coincides with the infinite-redshift
surface $r=0$. The alternative assumption $a-b+m r=0$ yields a
simpler set of algebraic equations, the solution of which,
however, does not satisfy the full set of Einstein's equations.

\section{Conclusions}

     We hope to have demonstrated here the power of our
nonholonomic approach to relativistic rotating matter. In the last
section, we have been able to carry out an exhaustive treatment of
incompressible perfect fluids with a vanishing complex tensor
$S_i^{\ell}$. Our investigation has lead to an unexpected conclusion:
the incompressible Schwarzschild space-time is characterized
with a vanishing Simon tensor. It is, however, not possible to set
it in rotation while retaining the zero value of the corresponding
complex tensor.
 
   There remains the freedom of adding a symmetric and trace-free
term to the complex tensor $S_i^{\ell}$ which vanishes in the static
and in the vacuum limits. The only expression with these
properties has the form $L_{(i}(G_{k)}-\bar G_{k)})T^*_{oo}$.
This term, however, contributes to the component $U_{\bf2}$, thus
in its presence the function ${\alpha}$ ceases to be analytic.
We intend to investigate rotating
fluids with this non-analytic property in the
sequel.

\section*{Acknowledgement}
 
We are grateful to J. Kadlecsik for supplying us with
a software package for noncommutative derivatives.
M.M. would like to thank Z.P. and the rest of his family for their
hospitality during the stay in Hungary.
M.M. would also like to thank Prof. Lennart Stenflo at the
Department of Plasma Physics, Ume{\aa} University, for financial
support.

\appendix
 
\section{Ricci tensor for axisymmetry}
 
  The inverse of the metric (\ref{eq:g^ik}) has the form
\begin{equation} \label{eq:g_ik}
 [{g_{\bf ik}}]= \left[ \matrix{
         C    &   -B   &   0 \cr
        -B    &    A   &   0 \cr
         0    &    0   &   \rho^{2} \cr } \right] , 
\end{equation}
with
\begin{equation}
  A=\frac{{\alpha}}{D}\ ,\qquad B=\frac{{\beta}}{D}
  \ ,\qquad C=\frac{{\gamma}}{D}\ .
\end{equation}
The nontrivial components of field equations (\ref{eq:Rij}) become
\begin{mathletters}
  \begin{eqnarray}
    R_{\bf11}&=&-\frac{k}{r^2}(T^*_{\bf11}-CT^*_{oo}) , \label{eq:r11} \\
    R_{\bf12}&=&-\frac{k}{r^2}(T^*_{\bf12}-BT^*_{oo})-\frac{1}{4r^2} ,
                                                  \label{eq:r12}\\
    R_{\bf22}&=&-\frac{k}{r^2}(T^*_{\bf22}-AT^*_{oo}) , \label{eq:r22} \\
    R_{\bf33}&=&-\frac{k}{r^2}(T^*_{\bf33}-\rho^2T^*_{oo}) . 
    \label{eq:r33} 
  \end{eqnarray}
\end{mathletters}
 
The components of the Ricci tensor for this metric are
\begin{mathletters}
\begin{eqnarray}
R_{\bf11}&=&  -(1/4)\{2 [(A + B) ({\alpha}{\gamma}
 -2{\beta}^2) {\varepsilon}
 + \rho_{,\bf1} {\alpha}/\rho + \rho_{,\bf2} {\beta}/\rho] C_{,\bf1}
 - 4 ({\gamma}_{,\bf1} {\varepsilon} + 
 {\varepsilon}_{,\bf1} {\gamma}) (A + B)
 + 8 {\beta} A B {\gamma} {\varepsilon}^2  \nonumber \\&&
+ 4[({\alpha} {\gamma} - {\beta}^2) C_{,\bf2}/2
 +  B_{,\bf2} {\beta} {\gamma}  -  {\gamma}_{,\bf2}
  -  \rho_{,\bf1} {\beta}/\rho -  \rho_{,\bf2} {\gamma}/\rho +
({\alpha} {\beta} - 2 {\alpha} {\gamma} + 4 {\beta}^2) C {\varepsilon}]
B_{,\bf1} \nonumber \\&&
 + [({\alpha} {\gamma} - 2 {\beta}^2) C_{,\bf1}
 + 2 ({\beta} - 2 {\gamma}) B {\gamma} {\varepsilon}
 + 2 B_{,\bf1} {\beta} {\gamma}
 - 2 {\gamma}_{,\bf1} - C_{,\bf2} {\beta} {\gamma} - 
 4 A {\gamma}^2 {\varepsilon} +
 2 {\beta} {\gamma} C {\varepsilon} - 4{\gamma}{\varepsilon}]A_{,\bf1}  
 \nonumber \\&&
 - (2 B_{,\bf1} {\gamma} - C_{,\bf1} {\beta} + C_{,\bf2} {\gamma} + 2 B
 {\gamma} {\varepsilon} + 2 {\gamma} C {\varepsilon}) A_{,\bf2} {\gamma}
 - 2 (C_{,\bf2} + 2 B {\varepsilon} + 2 C {\varepsilon}) {\gamma}_{,\bf2}
 + 4 (2 {\beta} - {\gamma}) B^2 {\gamma} {\varepsilon}^2   \nonumber \\&&
 - 2 (C_{,\bf1} {\beta}^2 - C_{,\bf2} {\beta} {\gamma} 
 - 2 B {\beta} {\gamma}
 {\varepsilon} - 2 {\beta} {\gamma} C {\varepsilon} + 
 2 {\gamma} {\varepsilon}) B_{,\bf2}
  - 4 ({\varepsilon}_{,\bf2} {\gamma}  + 
 \rho_{,\bf1} {\beta} {\varepsilon}/\rho + \rho_{,\bf2}
 {\gamma} {\varepsilon}/\rho) (B + C) \nonumber \\&&
 - 2 (\rho_{,\bf1} {\beta}/\rho + \rho_{,\bf2}
 {\gamma}/\rho + B {\beta}^2 {\varepsilon}  + {\beta}^2 C {\varepsilon}  
 - {\beta} {\varepsilon}
  + 2 {\gamma} {\varepsilon}  ) C_{,\bf2}
 + ({\alpha} {\gamma} - 2 {\beta}^2)(C_{,\bf2})^2
  - 4 (B + C)^2 {\alpha} {\gamma} {\varepsilon}^2
   \nonumber \\&&
 - 2 A_{,\bf11} {\gamma} - (A_{,\bf1})^2 {\gamma}^2
 - 4 B_{,\bf12} {\gamma}  - 2 {\beta}_{,\bf1} C_{,\bf2}
 + 2 {\beta}_{,\bf2} C_{,\bf1}  - 2 C_{,\bf22} {\gamma}
 - 4 \rho_{,\bf11}/\rho + 4(2B^2 - A^2) {\gamma}^2 {\varepsilon}^2 \}   ,
 \nonumber \\
 \nonumber  \\
R_{\bf12}&=& -(1/4)\{2 [({\beta} B + 
2{\beta} C - 1) {\alpha} {\varepsilon}  + {\alpha}
{\varepsilon}  + \rho_{,\bf1} {\alpha}/\rho + \rho_{,\bf2}
{\beta}/\rho - B {\beta}^2 {\varepsilon}  + 
2 {\beta} {\varepsilon}  ] C_{,\bf2}  \nonumber \\&&
 + 4 [{\beta} B {\alpha} {\varepsilon} - ({\alpha} {\gamma} + {\beta}^2)
A {\varepsilon} - B_{,\bf2} {\alpha} {\gamma} + {\beta}_{,\bf2} 
- B {\beta}^2
{\varepsilon} + {\beta} {\varepsilon}] B_{,\bf1}  \nonumber \\&&
 + 2 (B  +  A){\beta} C_{,\bf1} {\alpha} {\varepsilon}  - 2
 [2 (B + C) {\alpha} {\gamma} {\varepsilon} - C_{,\bf1} {\alpha} {\beta}
 + C_{,\bf2} {\alpha} {\gamma} - 2 {\beta} {\varepsilon}] B_{,\bf2}
 - 8 {\beta} B A {\beta} {\varepsilon}^2  \nonumber \\&&
 - [2 ({\beta} - {\gamma}) B {\beta} {\varepsilon}  + 
 2 B_{,\bf1} {\alpha} {\gamma}  - 2
 {\beta}_{,\bf1}  - C_{,\bf1} {\alpha} {\beta}  - C_{,\bf2} {\beta}
^2  - 2 \rho_{,\bf1} {\beta}/\rho - 2 \rho_{,\bf2} {\gamma}/\rho
 - 4 A{\beta} {\gamma} {\varepsilon} - 4 {\beta} {\varepsilon}  ]
A_{,\bf1} \nonumber \\&&
 + (2 B_{,\bf1} {\gamma} - C_{,\bf1} {\beta} + C_{,\bf2} {\gamma}
 + 2 B {\gamma} {\varepsilon} + 2{\gamma}C{\varepsilon})A_{,\bf2}{\beta}
 + 4 ({\beta}_{,\bf1} {\varepsilon}  + {\varepsilon}_{,\bf1} {\beta}
 + \rho_{,\bf1} {\beta} {\varepsilon}/\rho + 
 \rho_{,\bf2} {\gamma} {\varepsilon}/\rho)
 (A +B) \nonumber \\&&
 + 2 (C_{,\bf2} + 2 B {\varepsilon} + 2 C {\varepsilon}) {\beta}_{,\bf2}
 + 4 (B^2 + C^2) {\alpha} {\beta} {\varepsilon}^2
 + 4 (B + C) {\varepsilon}_{,\bf2}{\beta} - 8 B {\beta}
 {\varepsilon}^2  + 2 A_{,\bf11} {\beta}
 + (A_{,\bf1})^2 {\beta} {\gamma}  \nonumber \\&&
 + 2 {\alpha}_{,\bf1} C_{,\bf2}  - 2 {\alpha}_{,\bf2} C_{,\bf1}
 + 4 B_{,\bf12} {\beta}  + 2 C_{,\bf22}
 {\beta}  + (C_{,\bf2})^2 {\alpha} {\beta}  - 4 \rho_{,\bf12}/\rho + 4
 A^2 {\beta} {\gamma} {\varepsilon}^2  - 
 4 B {\beta}^2 C {\varepsilon}^2  \}  , \nonumber \\
 \nonumber \\
 R_{\bf22}&=& -(1/4)\{2 [(4 B + 2 C) {\alpha} {\beta} {\varepsilon}
 - C_{,\bf1} {\alpha}^2  + C_{,\bf2} {\alpha} {\beta}  - 
 2 \rho_{,\bf1} {\alpha}/\rho
 - 2 \rho_{,\bf2} {\beta}/\rho - 2 A {\alpha} {\gamma} {\varepsilon} + 
 ({\alpha} {\gamma} - {\beta}^2)
  A_{,\bf1}  ] B_{,\bf2} \nonumber \\&&
 +2 [
 -   {\alpha}_{,\bf1} - {\beta}_{,\bf2}
 + {\alpha} B_{,\bf1} {\beta}
 - {\alpha} (C_{,\bf1} {\alpha} + C_{,\bf2} {\beta})/2 -
(\rho_{,\bf1}{\alpha} + \rho_{,\bf2} {\beta})/\rho
 + (A - 2 B)({\beta}^2 + {\alpha}{\gamma}) {\varepsilon}
] A_{,\bf1}  \nonumber \\&&
 + [({\alpha} {\gamma} - 2 {\beta}^2) (C_{,\bf2} +2B{\varepsilon} + 
 2C{\varepsilon})
 - 2 B_{,\bf1} {\beta}^2  + 2 {\beta}_{,\bf1}  +
C_{,\bf1} {\alpha} {\beta}  + 2 \rho_{,\bf1} {\beta}/\rho + 2 \rho_{,\bf2}
 {\gamma}/\rho ] A_{,\bf2}\nonumber \\&&
 + 4 [(B {\beta} - 2) {\varepsilon} + B_{,\bf2} {\beta}
 + A {\beta} {\varepsilon}] B_{,\bf1} {\alpha}  
 - 2 {\beta} (B + A) C_{,\bf2}{\alpha} {\varepsilon}
 + 4 (2 {\beta} B + {\beta} C) {\alpha} B {\varepsilon}
^2  + 8 {\beta} B A {\alpha} {\varepsilon}^2  \nonumber \\&&
 - 4 (B_{,\bf2} + A {\varepsilon} + B {\varepsilon}) {\alpha}_{,\bf1}
 - 2 (C_{,\bf1} {\alpha}^2 {\varepsilon}  
 + 2 {\varepsilon}_{,\bf1} {\alpha}  + 2
 \rho_{,\bf1} {\alpha} {\varepsilon}/\rho + 
 2 \rho_{,\bf2} {\beta} {\varepsilon}/\rho)
 (A + B) \nonumber \\&&
- 2 (C_{,\bf2} + 2 B {\varepsilon} + 2 C {\varepsilon}) {\alpha}_{,\bf2}
 + ({\alpha} {\gamma} - 2 {\beta}^2) (A_{,\bf1})^2  - 4 (B + C)^2
 {\alpha}^2 {\varepsilon}^2   \nonumber \\&&
 - 4 (B + C) {\varepsilon}_{,\bf2} {\alpha}  - 2
A_{,\bf11} {\alpha}
 - 4 B_{,\bf12} {\alpha}
 - 2 C_{,\bf22} {\alpha}  - (C_{,\bf2})^2 {\alpha}^2  
 - 4 \rho_{,\bf22}/\rho
 - 4 A^2 {\alpha} {\gamma} {\varepsilon}^2 \}   , \nonumber \\
 \nonumber  \\
R_{\bf33}&=&[({\alpha} + 2 {\beta}) \rho_{,\bf1} {\varepsilon} + {\gamma}
 \rho_{,\bf2} {\varepsilon} + (A_{,\bf1} {\gamma} - 2 B_{,\bf1} {\beta}
 + C_{,\bf1} {\alpha}) (\rho_{,\bf1} {\alpha} + 
 \rho_{,\bf2} {\beta})/2\nonumber \\&&
 \qquad  \qquad  \qquad  \qquad  \ \, \quad
 + (A_{,\bf2} {\gamma} - 2 B_{,\bf2} {\beta} + C_{,\bf2} {\alpha})
 (\rho_{,\bf1} {\beta} + \rho_{,\bf2} {\gamma})/2\nonumber \\&&
 \qquad  \qquad  \qquad  \qquad  \ \, \quad
 + {\alpha}_{,\bf1} \rho_{,\bf1}
 + {\beta}_{,\bf1} \rho_{,\bf2} + {\beta}_{,\bf2} \rho_{,\bf1}
 + {\gamma}_{,\bf2} \rho_{,\bf2} + 2 \rho_{,\bf12} {\beta}
 + \rho_{,\bf11} {\alpha} + \rho_{,\bf22} {\gamma}] \rho , \nonumber
\end{eqnarray}
\end{mathletters}
and $R_{\bf13}=R_{\bf23}=0$.
 
\section{Perfect fluid equations}
  For an axisymmetric three-space, the Ricci tensor
has identically vanishing (1,3) and (2,3) components. Hence it
follows from the Einstein equations (\ref{eq:feq})
that the 4-velocity of the perfect fluid has the form
\[
(u_{\mu})=(u_o,0,0,u_{\varphi})\ .
\]
We shall also use the component $u^{\varphi}=u_{\varphi}/\rho^2$.
 
Einstein's equations for a perfect fluid are the condition that
the following expressions vanish:
\begin{mathletters}
  \begin{eqnarray}
    E_{00} & = & ({{\omega}}_{,A})^2 r^4 + (\rho r_{,A})_{,A} r {\rho} -
    (r_{,A})^2 {\rho}^2 - {e^{2{\lambda}}}k[ r {\rho}^2({\mu} + 3 p)
    + 2({\mu}+p){u_{{\varphi}}}^2] , \\
    E_{12} & = & {{\omega}}_{,1} {{\omega}}_{,2} r^4 
    + 2({\lambda}_{,1} {\rho}
    _{,2} + {\lambda}_{,2} {\rho}_{,1})r^2 {\rho} - r_{,1}
    r_{,2} {\rho}^2 - 2 {\rho}_{,12} r^2 {\rho} , \\ 
    E_{33} & = &r\rho(r^2\rho^{-1}{{\omega}}_{,A})_{,A}
    + 2 k {u_{{\varphi}}} u_o {e^{2{\lambda}}}({\mu}+p) , \\
    E_{1} & {\stackrel{\rm def}{=}} & E_{11} + E_{22} - 2r\rho E_3  
    \nonumber \\ 
    & = & ({\omega}_{,A})^2r^4 + 4{\lambda}_{,AA}r^2\rho^2 
    + (r_{,A})^2\rho^2
    - 4 {e^{2{\lambda}}}k[r {\rho}^2 p +
    ({\mu} + p){u_{{\varphi}}}^2]  , \\ 
    E_{2} & {\stackrel{\rm def}{=}} & E_{11} - E_{22}  \nonumber \\
    & = & [({{\omega}}_{,1})^2 - ({{\omega}}_{,2})^2]r^4 + 
    4[{\lambda}_{,1} {\rho}_{,1} -
    {\lambda}_{,2} {\rho}_{,2}]r^2 {\rho} \nonumber \\ 
    && - [(r_{,1})^2 - (r_{,2})^2]{\rho}^2 - 2[
    {\rho}_{,11} - {\rho}_{,22}]r^2 {\rho} , \\ 
    E_{3} & {\stackrel{\rm def}{=}} & E^0_0 + E^3_3 = {\rho}_{,AA} r - 
    2 {e^{2{\lambda}}}{\rho}
    kp , \\
    E_{4} & {\stackrel{\rm def}{=}} & E^0_0 - E^3_3  \nonumber \\ 
    & = & [r^2{\rho}^{-1}(\rho^2r^{-2} - {\omega}^2)_{,A}]_{,A}
    -2ke^{2{\lambda}} r^{-2}\rho^{-1}
    (\mu + p)(2r{\omega} u_o u_{{\varphi}} - r\rho^2 - 2u_{{\varphi}}^2) , 
  \end{eqnarray}
\end{mathletters}
where we assume summation over repeated indices. The equation
$E_4=0$ is a linear combination of the Eqs. for 
$E_{00},\ E_{33}$ and $E_3$.

\section{Coefficients in $Q_{{\alpha}}$}

Below we list the coefficients in the polynomial (\ref{eq:quad}),
given in Sec. \ref{sec:IntCond}:
\begin{eqnarray*}
a_1&=&
 2\,{\beta}^{5}k\mu
\,{r}^{2}{\psi}^{5}-3\,{\beta}^{4}{\alpha}\,r\gamma\,{\psi}^{6}-5\,{
\beta}^{4}{k}^{2}{\mu}^{2}{r}^{3}{\psi}^{4}-4\,{\beta}^{3}{\alpha}\,k
\mu\,{r}^{2}\gamma\,{\psi}^{5}-4\,{\beta}^{3}{k}^{3}{\mu}^{3}{r}^{4}
{\psi}^{3}+3\,{\beta}^{2}{{\alpha}}^{2}r{\gamma}^{2}{\psi}^{6}
\nonumber\\&&
+10\,{
\beta}^{2}{\alpha}\,{k}^{2}{\mu}^{2}{r}^{3}\gamma\,{\psi}^{4}+3\,{
\beta}^{2}{k}^{4}{\mu}^{4}{r}^{5}{\psi}^{2}+2\,\beta\,{{\alpha}}^{2}k
\mu\,{r}^{2}{\gamma}^{2}{\psi}^{5}+4\,\beta\,{\alpha}\,{k}^{3}{\mu}^{3
}{r}^{4}\gamma\,{\psi}^{3}-6\,\beta\,{k}^{5}{\mu}^{5}{r}^{6}\psi
\nonumber\\&&
-5\,
{{\alpha}}^{2}{k}^{2}{\mu}^{2}{r}^{3}{\gamma}^{2}{\psi}^{4}-3\,{\alpha}
\,{k}^{4}{\mu}^{4}{r}^{5}\gamma\,{\psi}^{2}+9\,{k}^{6}{\mu}^{6}{r}^{
7}+{\beta}^{6}r{\psi}^{6}-{{\alpha}}^{3}r{\gamma}^{3}{\psi}^{6} \ \ , \\
b_1&=&
   -{
{\alpha}}^{3}{\gamma}^{3}{\psi}^{6}-8\,{\beta}^{4}{k}^{2}{\mu}^{2}{r}^
{2}{\psi}^{5}-3\,{\beta}^{4}{k}^{2}{\mu}^{2}{r}^{2}{\psi}^{4}-8\,{
\beta}^{3}{k}^{3}{\mu}^{3}{r}^{3}{\psi}^{4}-6\,{\beta}^{3}{k}^{3}{
\mu}^{3}{r}^{3}{\psi}^{3}+2\,{\beta}^{2}{{\alpha}}^{2}{\gamma}^{2}{
\psi}^{6}
\nonumber\\&&
+16\,{\beta}^{2}{\alpha}\,{k}^{2}{\mu}^{2}{r}^{2}\gamma\,{
\psi}^{5}+8\,{\beta}^{2}{\alpha}\,{k}^{2}{\mu}^{2}{r}^{2}\gamma\,{\psi
}^{4}+16\,{\beta}^{2}{k}^{4}{\mu}^{4}{r}^{4}{\psi}^{3}+6\,{\beta}^{2
}{k}^{4}{\mu}^{4}{r}^{4}{\psi}^{2}+8\,\beta\,{\alpha}\,{k}^{3}{\mu}^{3
}{r}^{3}\gamma\,{\psi}^{4}
\nonumber\\&&
+6\,\beta\,{\alpha}\,{k}^{3}{\mu}^{3}{r}^{3}
\gamma\,{\psi}^{3}-8\,\beta\,{k}^{5}{\mu}^{5}{r}^{5}{\psi}^{2}+18\,
\beta\,{k}^{5}{\mu}^{5}{r}^{5}\psi-8\,{{\alpha}}^{2}{k}^{2}{\mu}^{2}{r
}^{2}{\gamma}^{2}{\psi}^{5}-5\,{{\alpha}}^{2}{k}^{2}{\mu}^{2}{r}^{2}{
\gamma}^{2}{\psi}^{4}
\nonumber\\&&
-16\,{\alpha}\,{k}^{4}{\mu}^{4}{r}^{4}\gamma\,{
\psi}^{3}-11\,{\alpha}\,{k}^{4}{\mu}^{4}{r}^{4}\gamma\,{\psi}^{2}+24\,
{k}^{6}{\mu}^{6}{r}^{6}\psi+9\,{k}^{6}{\mu}^{6}{r}^{6}-{\beta}^{4}
{\alpha}\,\gamma\,{\psi}^{6} \ \ , \\
c_1&=&
  4\,{\beta}^{2}{\alpha}\,{k}^{2}{\mu}^{2}r
\gamma\,{\psi}^{5}+4\,{\beta}^{2}{\alpha}\,{k}^{2}{\mu}^{2}r\gamma\,{
\psi}^{4}+16\,{\beta}^{2}{k}^{4}{\mu}^{4}{r}^{3}{\psi}^{4}+12\,{
\beta}^{2}{k}^{4}{\mu}^{4}{r}^{3}{\psi}^{3}+4\,\beta\,{\alpha}\,{k}^{3
}{\mu}^{3}{r}^{2}\gamma\,{\psi}^{3}
\nonumber\\&&
+24\,\beta\,{k}^{5}{\mu}^{5}{r}^{
4}{\psi}^{2}-4\,{{\alpha}}^{2}{k}^{2}{\mu}^{2}r{\gamma}^{2}{\psi}^{5}-
4\,{{\alpha}}^{2}{k}^{2}{\mu}^{2}r{\gamma}^{2}{\psi}^{4}-16\,{\alpha}\,{
k}^{4}{\mu}^{4}{r}^{3}\gamma\,{\psi}^{4}-8\,{\alpha}\,{k}^{4}{\mu}^{4}
{r}^{3}\gamma\,{\psi}^{3}
\nonumber\\&&
+16\,{k}^{6}{\mu}^{6}{r}^{5}{\psi}^{2}+12\,
{k}^{6}{\mu}^{6}{r}^{5}\psi \ \ ,  \\
\nonumber \\
a_2&=&
72\,\beta\,{\alpha}\,{k}^{4}{\mu}^{4}{r}^{6}\gamma\,{
\psi}^{2}
 -92\,{\beta}^{2}{{\alpha}}
^{2}k\mu\,{r}^{3}{\gamma}^{2}{\psi}^{5}-160\,{\beta}^{2}{\alpha}\,{k}^
{3}{\mu}^{3}{r}^{5}\gamma\,{\psi}^{3}-36\,{\beta}^{2}{k}^{5}{\mu}^{5
}{r}^{7}\psi  
\\ && 
+24\,\beta\,{{\alpha}}^{2}{k}^{2}{\mu}^{2}{r}^{4}{\gamma}^{2}{\psi}^{4}
+28\,{{\alpha}}^{3}k\mu\,{r}^{3}{\gamma}^{3}{\psi}^{5}+40\,{
{\alpha}}^{2}{k}^{3}{\mu}^{3}{r}^{5}{\gamma}^{2}{\psi}^{3}+60\,{\alpha}
\,{k}^{5}{\mu}^{5}{r}^{7}\gamma\,\psi 
\\ &&
- 36\,{\beta}^{6}k\mu\,{r}^{3}{\psi}^{5} + 
24\,{\beta}^{5}{k}^{2}{\mu}^{2}{r}^{4}{\psi}^{4}
+100\,{
\beta}^{4}{\alpha}\,k\mu\,{r}^{3}\gamma\,{\psi}^{5}+120\,{\beta}^{4}{k
}^{3}{\mu}^{3}{r}^{5}{\psi}^{3} 
\\ &&
- 48\,{\beta}^{3}{\alpha}\,{k}^{2}{\mu}^
{2}{r}^{4}\gamma\,{\psi}^{4}-72\,{\beta}^{3}{k}^{4}{\mu}^{4}{r}^{6}{
\psi}^{2} \ \ , \\
b_2&=&
        -46\,{\beta}^{3}{\alpha}\,{k}^{2}{\mu}^{2}{r}^{3}\gamma\,{\psi
}^{4}-96\,{\beta}^{3}{k}^{4}{\mu}^{4}{r}^{5}{\psi}^{3}-15\,{\beta}^{
3}{k}^{4}{\mu}^{4}{r}^{5}{\psi}^{2}+64\,{\alpha}\,{k}^{5}{\mu}^{5}{r}^
{6}\gamma\,\psi+24\,{\beta}^{4}{\alpha}\,k\mu\,{r}^{2}\gamma\,{\psi}^{
5}
\nonumber\\&&
+192\,{\beta}^{4}{k}^{3}{\mu}^{3}{r}^{4}{\psi}^{4}+36\,{\beta}^{4}
{k}^{3}{\mu}^{3}{r}^{4}{\psi}^{3}+{\beta}^{3}{{\alpha}}^{2}r{\gamma}^{
2}{\psi}^{6}+\beta\,{{\alpha}}^{3}r{\gamma}^{3}{\psi}^{6}+3\,{\beta}^{
7}r{\psi}^{6}-48\,{\beta}^{2}{{\alpha}}^{2}k\mu\,{r}^{2}{\gamma}^{2}{
\psi}^{5}
\nonumber\\&&
-352\,{\beta}^{2}{\alpha}\,{k}^{3}{\mu}^{3}{r}^{4}\gamma\,{
\psi}^{4}-192\,{\beta}^{2}{k}^{5}{\mu}^{5}{r}^{6}{\psi}^{2}-124\,{
\beta}^{2}{\alpha}\,{k}^{3}{\mu}^{3}{r}^{4}\gamma\,{\psi}^{3}-108\,{
\beta}^{2}{k}^{5}{\mu}^{5}{r}^{6}\psi
\nonumber\\&&
+43\,\beta\,{{\alpha}}^{2}{k}^{2}
{\mu}^{2}{r}^{3}{\gamma}^{2}{\psi}^{4}+24\,{{\alpha}}^{3}k\mu\,{r}^{2}
{\gamma}^{3}{\psi}^{5}+96\,\beta\,{\alpha}\,{k}^{4}{\mu}^{4}{r}^{5}
\gamma\,{\psi}^{3}+3\,\beta\,{\alpha}\,{k}^{4}{\mu}^{4}{r}^{5}\gamma\,
{\psi}^{2}-63\,\beta\,{k}^{6}{\mu}^{6}{r}^{7}
\nonumber\\&&
-5\,{\beta}^{5}{\alpha}\,
r\gamma\,{\psi}^{6}+3\,{\beta}^{5}{k}^{2}{\mu}^{2}{r}^{3}{\psi}^{4}+
160\,{{\alpha}}^{2}{k}^{3}{\mu}^{3}{r}^{4}{\gamma}^{2}{\psi}^{4}+88\,{
{\alpha}}^{2}{k}^{3}{\mu}^{3}{r}^{4}{\gamma}^{2}{\psi}^{3}+224\,{\alpha}
\,{k}^{5}{\mu}^{5}{r}^{6}\gamma\,{\psi}^{2}  \ \ , \\
c_2 &=&
-4\,{\beta}^{2}{{\alpha}}^{2}k\mu\,r{\gamma}^{2}{\psi}^{5
}-80\,{\beta}^{2}{\alpha}\,{k}^{3}{\mu}^{3}{r}^{3}\gamma\,{\psi}^{4}-
64\,{\beta}^{2}{\alpha}\,{k}^{3}{\mu}^{3}{r}^{3}\gamma\,{\psi}^{3}-192
\,{\beta}^{2}{k}^{5}{\mu}^{5}{r}^{5}{\psi}^{3} 
\\ &&
- 144\,{\beta}^{2}{k}^{5}{\mu}^{5}{r}^{5}{\psi}^{2}
-2\,\beta\,{{\alpha}}^{3}{\gamma}^{3}{\psi
}^{6}+4\,\beta\,{{\alpha}}^{2}{k}^{2}{\mu}^{2}{r}^{2}{\gamma}^{2}{\psi
}^{5}+32\,\beta\,{{\alpha}}^{2}{k}^{2}{\mu}^{2}{r}^{2}{\gamma}^{2}{
\psi}^{4} 
\\ &&
+ 32\,\beta\,{\alpha}\,{k}^{4}{\mu}^{4}{r}^{4}\gamma\,{\psi}^{
3}-38\,\beta\,{\alpha}\,{k}^{4}{\mu}^{4}{r}^{4}\gamma\,{\psi}^{2}
-84\,
\beta\,{k}^{6}{\mu}^{6}{r}^{6}\psi+2\,{{\alpha}}^{3}k\mu\,r{\gamma}^{3
}{\psi}^{5}+80\,{{\alpha}}^{2}{k}^{3}{\mu}^{3}{r}^{3}{\gamma}^{2}{\psi
}^{4} 
\\ &&
+ 64\,{{\alpha}}^{2}{k}^{3}{\mu}^{3}{r}^{3}{\gamma}^{2}{\psi}^{3}+
192\,{\alpha}\,{k}^{5}{\mu}^{5}{r}^{5}\gamma\,{\psi}^{3}
+112\,{\alpha}\,
{k}^{5}{\mu}^{5}{r}^{5}\gamma\,{\psi}^{2}-10\,{\alpha}\,{k}^{5}{\mu}^{
5}{r}^{5}\gamma\,\psi-12\,{\beta}^{5}{k}^{2}{\mu}^{2}{r}^{2}{\psi}^{
5} 
\\ &&
- 2\,{\beta}^{5}{\alpha}\,\gamma\,{\psi}^{6}+2\,{\beta}^{4}{\alpha}\,k
\mu\,r\gamma\,{\psi}^{5}
+4\,{\beta}^{3}{{\alpha}}^{2}{\gamma}^{2}{\psi
}^{6}+8\,{\beta}^{3}{\alpha}\,{k}^{2}{\mu}^{2}{r}^{2}\gamma\,{\psi}^{5
} 
\\ &&
- 32\,{\beta}^{3}{\alpha}\,{k}^{2}{\mu}^{2}{r}^{2}\gamma\,{\psi}^{4}-
48\,{\beta}^{3}{k}^{4}{\mu}^{4}{r}^{4}{\psi}^{3} \ \ , \\
\ \\
a_3&=&
   -24\,{\beta}^{3}{k}^{4}{
\mu}^{4}{r}^{6}+8\,{\beta}^{7}{r}^{2}{\psi}^{4}+16\,{\beta}^{6}k\mu
\,{r}^{3}{\psi}^{3}-24\,{\beta}^{5}{\alpha}\,{r}^{2}\gamma\,{\psi}^{4}
-16\,{\beta}^{5}{k}^{2}{\mu}^{2}{r}^{4}{\psi}^{2} 
\\ &&
- 48\,{\beta}^{4}
{\alpha}\,k\mu\,{r}^{3}\gamma\,{\psi}^{3}
+16\,{\beta}^{4}{k}^{3}{\mu}^
{3}{r}^{5}\psi+24\,{\beta}^{3}{{\alpha}}^{2}{r}^{2}{\gamma}^{2}{\psi}^
{4}+32\,{\beta}^{3}{\alpha}\,{k}^{2}{\mu}^{2}{r}^{4}\gamma\,{\psi}^{2}
\\ &&
+ 48\,{\beta}^{2}{{\alpha}}^{2}k\mu\,{r}^{3}{\gamma}^{2}{\psi}^{3}-32\,
{\beta}^{2}{\alpha}\,{k}^{3}{\mu}^{3}{r}^{5}\gamma\,\psi
-8\,\beta\,{
{\alpha}}^{3}{r}^{2}{\gamma}^{3}{\psi}^{4}-16\,\beta\,{{\alpha}}^{2}{k}^
{2}{\mu}^{2}{r}^{4}{\gamma}^{2}{\psi}^{2} 
\\ &&
+24\,\beta\,{\alpha}\,{k}^{4}
{\mu}^{4}{r}^{6}\gamma-16\,{{\alpha}}^{3}k\mu\,{r}^{3}{\gamma}^{3}{
\psi}^{3}+16\,{{\alpha}}^{2}{k}^{3}{\mu}^{3}{r}^{5}{\gamma}^{2}\psi \ , \\
b_3&=&
40\,{\beta}^{3}{{\alpha}}^{2}r{\gamma}^{2}{\psi}^{4}\!
-20\,{\beta}^{5}{\alpha}\,r\gamma\,{\psi}^{4}-32\,{\beta}^{5}{k}^{2}{
\mu}^{2}{r}^{3}{\psi}^{3}-24\,{\beta}^{5}{k}^{2}{\mu}^{2}{r}^{3}{
\psi}^{2}-48\,{\beta}^{4}{k}^{3}{\mu}^{3}{r}^{4}\psi\!
\\ &&
-16\,{{\alpha}}^{2}{k}^{3}{\mu}^{3}{r}^{4}{\gamma}^{2}\psi
+64\,{\beta}^{3}{\alpha}\,{k}^{2}{
\mu}^{2}{r}^{3}\gamma\,{\psi}^{3}+40\,{\beta}^{3}{\alpha}\,{k}^{2}{\mu
}^{2}{r}^{3}\gamma\,{\psi}^{2}-32\,{\beta}^{3}{k}^{4}{\mu}^{4}{r}^{5
}\psi 
\\ &&
-24\,{\beta}^{3}{k}^{4}{\mu}^{4}{r}^{5}+64\,{\beta}^{2}{\alpha}\,
{k}^{3}{\mu}^{3}{r}^{4}\gamma\,\psi
-20\,\beta\,{{\alpha}}^{3}r{\gamma}
^{3}{\psi}^{4}-32\,\beta\,{{\alpha}}^{2}{k}^{2}{\mu}^{2}{r}^{3}{\gamma
}^{2}{\psi}^{3} \\ &&
-16\,\beta\,{{\alpha}}^{2}{k}^{2}{\mu}^{2}{r}^{3}{
\gamma}^{2}{\psi}^{2}+32\,\beta\,{\alpha}\,{k}^{4}{\mu}^{4}{r}^{5}
\gamma\,\psi+36\,\beta\,{\alpha}\,{k}^{4}{\mu}^{4}{r}^{5}\gamma \ \ , \\
c_3&=&
-7\,{\beta}^{5}
{\alpha}\,\gamma\,{\psi}^{4}+8\,{\beta}^{4}{\alpha}\,k\mu\,r\gamma\,{
\psi}^{3}+14\,{\beta}^{3}{{\alpha}}^{2}{\gamma}^{2}{\psi}^{4}+16\,{
\beta}^{3}{\alpha}\,{k}^{2}{\mu}^{2}{r}^{2}\gamma\,{\psi}^{3}+34\,{
\beta}^{3}{\alpha}\,{k}^{2}{\mu}^{2}{r}^{2}\gamma\,{\psi}^{2}
\nonumber\\&&
-8\,{
\beta}^{2}{{\alpha}}^{2}k\mu\,r{\gamma}^{2}{\psi}^{3}+24\,{\beta}^{2}
{\alpha}\,{k}^{3}{\mu}^{3}{r}^{3}\gamma\,\psi-7\,\beta\,{{\alpha}}^{3}{
\gamma}^{3}{\psi}^{4}-16\,\beta\,{{\alpha}}^{2}{k}^{2}{\mu}^{2}{r}^{2}
{\gamma}^{2}{\psi}^{3} 
\\ &&
-34\,\beta\,{{\alpha}}^{2}{k}^{2}{\mu}^{2}{r}^{2
}{\gamma}^{2}{\psi}^{2}
+16\,\beta\,{\alpha}\,{k}^{4}{\mu}^{4}{r}^{4}
\gamma\,\psi+5\,\beta\,{\alpha}\,{k}^{4}{\mu}^{4}{r}^{4}\gamma \ \ , \\
\  \\
a_4&=&
   8\,{\beta}^{7}r{\psi}^{6}-36\,{\beta}^{6}k\mu\,{r}^{2}{
\psi}^{5}-24\,{\beta}^{5}{\alpha}\,r\gamma\,{\psi}^{6}+120\,{\beta}^{5
}{k}^{2}{\mu}^{2}{r}^{3}{\psi}^{4}+100\,{\beta}^{4}{\alpha}\,k\mu\,{r}
^{2}\gamma\,{\psi}^{5} 
\\ &&
+72\,{\beta}^{4}{k}^{3}{\mu}^{3}{r}^{4}{\psi}^{3}
+24\,{\beta}^{3}{{\alpha}}^{2}r{\gamma}^{2}{\psi}^{6}-240\,{\beta}^
{3}{\alpha}\,{k}^{2}{\mu}^{2}{r}^{3}\gamma\,{\psi}^{4}-176\,{\beta}^{3
}{k}^{4}{\mu}^{4}{r}^{5}{\psi}^{2} 
\\ &&
-92\,{\beta}^{2}{{\alpha}}^{2}k\mu\,
{r}^{2}{\gamma}^{2}{\psi}^{5}
-192\,{\beta}^{2}{\alpha}\,{k}^{3}{\mu}^{
3}{r}^{4}\gamma\,{\psi}^{3}+12\,{\beta}^{2}{k}^{5}{\mu}^{5}{r}^{6}
\psi-8\,\beta\,{{\alpha}}^{3}r{\gamma}^{3}{\psi}^{6} 
\\ &&
+ 120\,\beta\,{
{\alpha}}^{2}{k}^{2}{\mu}^{2}{r}^{3}{\gamma}^{2}{\psi}^{4}+176\,\beta
\,{\alpha}\,{k}^{4}{\mu}^{4}{r}^{5}\gamma\,{\psi}^{2}
+28\,{{\alpha}}^{3}
k\mu\,{r}^{2}{\gamma}^{3}{\psi}^{5} 
\\ && 
+120\,{{\alpha}}^{2}{k}^{3}{\mu}^{3
}{r}^{4}{\gamma}^{2}{\psi}^{3}+12\,{\alpha}\,{k}^{5}{\mu}^{5}{r}^{6}
\gamma\,\psi \ \ ,  \\
b_4&=&
3\,{
\beta}^{7}{\psi}^{6}-144\,{\beta}^{2}{k}^{5}{\mu}^{5}{r}^{5}\psi-63
\,\beta\,{k}^{6}{\mu}^{6}{r}^{6}-12\,{\beta}^{6}k\mu\,r{\psi}^{5}+
256\,{\beta}^{4}{k}^{3}{\mu}^{3}{r}^{3}{\psi}^{4}-21\,{\beta}^{5}{k}
^{2}{\mu}^{2}{r}^{2}{\psi}^{4}
\nonumber\\&&
-12\,{\beta}^{4}{k}^{3}{\mu}^{3}{r}^{3
}{\psi}^{3}-87\,{\beta}^{3}{k}^{4}{\mu}^{4}{r}^{4}{\psi}^{2}-4\,{
\beta}^{2}{{\alpha}}^{2}k\mu\,r{\gamma}^{2}{\psi}^{5}-480\,{\beta}^{2}
{\alpha}\,{k}^{3}{\mu}^{3}{r}^{3}\gamma\,{\psi}^{4} 
\\ &&
-92\,{\beta}^{2}
{\alpha}\,{k}^{3}{\mu}^{3}{r}^{3}\gamma\,{\psi}^{3}
-128\,{\beta}^{2}{k
}^{5}{\mu}^{5}{r}^{5}{\psi}^{2}+17\,\beta\,{{\alpha}}^{3}{\gamma}^{3}{
\psi}^{6}-32\,\beta\,{{\alpha}}^{2}{k}^{2}{\mu}^{2}{r}^{2}{\gamma}^{2}
{\psi}^{5} 
\\ &&
+75\,\beta\,{{\alpha}}^{2}{k}^{2}{\mu}^{2}{r}^{2}{\gamma}^{2
}{\psi}^{4}+320\,\beta\,{\alpha}\,{k}^{4}{\mu}^{4}{r}^{4}\gamma\,{\psi
}^{3}
+115\,\beta\,{\alpha}\,{k}^{4}{\mu}^{4}{r}^{4}\gamma\,{\psi}^{2}+
28\,{{\alpha}}^{3}k\mu\,r{\gamma}^{3}{\psi}^{5} 
\\ &&
+224\,{{\alpha}}^{2}{k}^{
3}{\mu}^{3}{r}^{3}{\gamma}^{2}{\psi}^{4}+104\,{{\alpha}}^{2}{k}^{3}{
\mu}^{3}{r}^{3}{\gamma}^{2}{\psi}^{3}+160\,{\alpha}\,{k}^{5}{\mu}^{5}{
r}^{5}\gamma\,{\psi}^{2}
+44\,{\alpha}\,{k}^{5}{\mu}^{5}{r}^{5}\gamma\,
\psi \\ &&
-32\,{\beta}^{5}{k}^{2}{\mu}^{2}{r}^{2}{\psi}^{5}+11\,{\beta}^{5
}{\alpha}\,\gamma\,{\psi}^{6}-12\,{\beta}^{4}{\alpha}\,k\mu\,r\gamma\,{
\psi}^{5}-31\,{\beta}^{3}{{\alpha}}^{2}{\gamma}^{2}{\psi}^{6}
\nonumber\\&&
+64\,{
\beta}^{3}{\alpha}\,{k}^{2}{\mu}^{2}{r}^{2}\gamma\,{\psi}^{5}-54\,{
\beta}^{3}{\alpha}\,{k}^{2}{\mu}^{2}{r}^{2}\gamma\,{\psi}^{4}-320\,{
\beta}^{3}{k}^{4}{\mu}^{4}{r}^{4}{\psi}^{3} \ \ , \\
c_4&=&
-12\,{\beta}^{5}{k}^{2}{\mu}^{2}r{\psi}^
{5}-14\,{\beta}^{4}{\alpha}\,k\mu\,\gamma\,{\psi}^{5}-120\,{\beta}^{3}
{\alpha}\,{k}^{2}{\mu}^{2}r\gamma\,{\psi}^{5}-80\,{\beta}^{3}{\alpha}\,{
k}^{2}{\mu}^{2}r\gamma\,{\psi}^{4}-48\,{\beta}^{3}{k}^{4}{\mu}^{4}{r
}^{3}{\psi}^{3}
\nonumber\\&&
+12\,{\beta}^{2}{{\alpha}}^{2}k\mu\,{\gamma}^{2}{\psi}^{5}\!
-80\,{\beta}^{2}{\alpha}\,{k}^{3}{\mu}^{3}{r}^{2}\gamma\,{\psi}^{4}
-104\,{\beta}^{2}{\alpha}\,{k}^{3}{\mu}^{3}{r}^{2}\gamma\,{\psi}^{3}-
192\,{\beta}^{2}{k}^{5}{\mu}^{5}{r}^{4}{\psi}^{3} 
\\ &&
-144\,{\beta}^{2}{k
}^{5}{\mu}^{5}{r}^{4}{\psi}^{2}
+132\,\beta\,{{\alpha}}^{2}{k}^{2}{\mu}
^{2}r{\gamma}^{2}{\psi}^{5}+112\,\beta\,{{\alpha}}^{2}{k}^{2}{\mu}^{2}
r{\gamma}^{2}{\psi}^{4}-48\,\beta\,{\alpha}\,{k}^{4}{\mu}^{4}{r}^{3}
\gamma\,{\psi}^{2} 
\\ &&
-84\,\beta\,{k}^{6}{\mu}^{6}{r}^{5}\psi+2\,{{\alpha}
}^{3}k\mu\,{\gamma}^{3}{\psi}^{5}
+80\,{{\alpha}}^{2}{k}^{3}{\mu}^{3}{r
}^{2}{\gamma}^{2}{\psi}^{4}+64\,{{\alpha}}^{2}{k}^{3}{\mu}^{3}{r}^{2}{
\gamma}^{2}{\psi}^{3} 
\\ &&
+192\,{\alpha}\,{k}^{5}{\mu}^{5}{r}^{4}\gamma\,{
\psi}^{3}+112\,{\alpha}\,{k}^{5}{\mu}^{5}{r}^{4}\gamma\,{\psi}^{2}-10
\,{\alpha}\,{k}^{5}{\mu}^{5}{r}^{4}\gamma\,\psi \ \ .
\end{eqnarray*}

\end{document}